\documentclass[11pt]{revtex4}
 \pdfoutput=1 
\usepackage{graphicx}
\begin{document}
\title { Determination of the second virial coefficient of interacting Bosons using the method of Wigner distribution function}
\author{Anirban Bose}
\affiliation {Serampore College, Serampore, Hooghly, India.\\}

\begin{abstract}
In a previous article \cite{kn:anirban1} a method has been introduced to derive the all order Bose-Einstein distribution of the non interacting Bosons as the  solution of the Wigner equation. The process was a perturbative one where the Bose-Einstein distribution was taken as the unperturbed solution. In this article it is shown that the same formalism is also applicable in the case of interacting Bosons. The formalism has been applied to calculate the quantum second virial coefficient of the Bosons interacting pairwise via Lenard-Jones potential and compared with the previous result.
\end{abstract}
% \pacs{05.30.-d, 05.20.Dd}
\maketitle

\newpage
\section{Introduction}

In a previous article  \cite{kn:anirban1} we have proposed a formalism to determine  the single particle distribution function as the all order solution of the Wigner equation in the case of non interacting Bosons placed in an external potential. The structure of the formalism suggests that the same technique may also be useful to determine the distribution function of Fermions as well.
In addition to that, it was observed that as we approach the low temperature and high density limit the importance of the exchange effect can not be neglected and in that limit the Bose-Einstein distribution may be the more appropriate choice as the zeroth order solution of the Wigner equation instead of the Maxwell-Boltzmann distribution.

In this article we have shown that the formalism introduced in \cite{kn:anirban1} to deal with the Bosons in the non interacting case may be further applied even to the systems of interacting Bosons and that actually extends the domain of application of the formalism.

In order to verify the range of application of the formalism we have chosen a specific problem of determination of virial coefficient of a system of interacting Bosons by using the same formalism  and compared it with the results already present in the literature. In fact, virial coefficients are found in a many body system in the virial expansion of pressure \cite{kn:path}. The expansion is expressed in the power of density of the system and that incorporates the corrections to the ideal system. The coefficients depend on the temperature as well as the interaction potential of the particles of the many body system. To be specific the second virial coefficient depends on the pairwise interaction potential of the particles.  
  
In the past various attempts have been made to introduce quantum correction to the virial coefficients. The first quantum correction to the classical value of the second virial coefficient for the square-well potential was calculated by T. S. Nilsen \cite{kn:ts}. The quantum formulae for the second virial coefficients is expressed in terms of the phases of the Schr\"{o}dinger wave function by L. Gropper\cite{kn:grop}. T. Kihara  has discussed the second virial coeffcient of helium for which the quantum effect is important \cite{kn:hara}. In the high temperature limit the second virial coefficient is obtained and the first correction term in the case of helium molecules interacting via Lenard-Jones potential is calculated by Uhlenbeck and Beth\cite{kn:uhl}. 
They have also considered the Bose and Fermi statistics in the calculation of the virial coefficient\cite{kn:uhl1}. Virial expansions (at fixed temperature) of the Maxwell-Boltzmann thermodynamic functions for a quantum plasma has been performed where the correlations have been represented by a diagrammatic series for a quantum plasma by Alastuey et al.\cite{kn:ala1,kn:ala2}. In the next step the exchange contributions due to Fermi or Bose statistics have been taken care of. The whole scheme is based on the Feynman-Kac path integral representation \cite{kn:ala3}. 
A numerical methodology has been developed to calculate quantum corrected virial coefficients using the centroid approximation of the exact path-integral expression \cite{kn:garb}.

Therefore, there is ample scope for comparison of our result with the previous results already available in the literature. 
\section{Determination of the quantum corrected second virial coefficient}
For this purpose, we consider a Bose gas confined in a volume V at temperature $T$. It is known that if the interaction between the molecules are neglected the ideal gas law is obtained. To determine the deviation from the ideal gas law we need to include the interactions of the molecules in the calculation. Therefore, to calculate the second virial coefficient we have considered the molecules are interacting pair wise via 
Lenard-Jones potential. We have actually restricted our calculations to binary encounters only. Therefore, the encounters involving more than two particles are neglected in the sufficiently low density limit.

The detailed discussion of obtaining the the virial coefficient may be found else where \cite{kn:ursell} and we do not need to repeat the same in this article.  The value of the second virial coefficient is given by \cite{kn:ursell} 
\begin{equation}B=-2\pi N\int_{_{0}}^{\infty}(S(r)-1)r^{2}dr\label{c6}\end{equation}
N is the number of molecules and $S(r)$ is the probability of finding  two molecules a distance r apart from each other. 

Now, we may employ our formalism to determine $S$.
In the semiclassical limit, the corrections may be separated in two parts. The first  part represents the direct quantum correction to the Boltzmann gas due to the inclusion of the terms containing the higher derivatives of the potential function in the Wigner equation in addition to the usual classical term as observed in the classical Boltzmann equation.  The second part arises because of the effects of the departure of a Bose-Einstein gas from Boltzmann statistics and that is commonly known as exchange effect. These two terms become more appreciable as the temperature of the system is diminished.

The quantum mechanical generalization of the Boltzmann equation is available in the literature \cite{kn:green, kn:imam} and the equation of the two particle distribution function of a system of molecules is given by \cite{kn:green}
\begin{eqnarray}
&&\frac{\partial f_{2}}{\partial t}+{\bf{v_{1}}}\cdot
\frac{\partial f_{2}}{\partial {\bf{r_{1}}}}+{\bf{v_{2}}}\cdot
\frac{\partial f_{2}}{\partial {\bf{r_{2}}}}=O^{(12)}f_{2}+\int \int (O^{(13)}+O^{(23)})f_{3}d\bf{r_{3}}d\bf{v_{3}}\label{v111}\end{eqnarray}

where the collision operator
\begin{eqnarray}
&&O^{(ij)}=\frac{2}{\hbar} \int d{\bf{k}}\chi({\bf{k}})\sin(m{\bf{k}}\cdot {\bf{r}}^{(ij)}/\hbar)\exp[\frac{{\bf{k}}}{2}\cdot(\frac{\partial }{\partial {\bf{v_{i}}}}-\frac{\partial }{\partial {\bf{v_{j}}}})]\label{v112}\end{eqnarray}
${\bf{r}}^{(ij)}={\bf{r_{j}}}-{\bf{r_{i}}}$ and $\chi({\bf{k}})=(\frac{m}{h})^3 \int d{\bf{r}}\phi({\bf{r}})\cos(m{\bf{k}}\cdot {\bf{r}}/\hbar)$

$\bf{r_{i}}$ and $\bf{r_{j}}$ are the position vectors of the i-th and j-th particle respectively. $\bf{v_{i}}$ and $\bf{v_{j}}$ are the velocity vectors of the i-th and j-th particle respectively.

$f_N$ is the N particle distribution function which is a function of time and position and velocity vectors of those N particles. 
$f_2$ and $f_3$ are the two and three particle distribution functions respectively.  $\phi$ is the two particle interaction potential. $\bf{r_{1}}$, $\bf{r_{2}}$, $\bf{r_{3}}$ and $\bf{v_{1}}$, $\bf{v_{2}}$, $\bf{v_{3}}$ are the position and velocity vectors of particle 1, 2 and 3 respectively. The mass of a single particle is $m$.

We may consider a pair of particles interacting with each other. The assumption of binary encounter \cite{kn:green} is based on the smallness of the probability of finding a third molecule in the close proximity of the aforementioned pair of molecules. This assumption is justified as long as the density of the system is sufficiently low. Therefore, in case of binary encounter the $f_{3}$ in eq.(\ref{v111}) may be safely neglected to obtain
\begin{eqnarray}
&&\frac{\partial f_{2}}{\partial t}+{\bf{v_{1}}}\cdot
\frac{\partial f_{2}}{\partial {\bf{r_{1}}}}+{\bf{v_{2}}}\cdot
\frac{\partial f_{2}}{\partial {\bf{r_{2}}}}=O^{(12)}f_{2}\label{v113}\end{eqnarray} 
Expanding the exponent of $O^{(12)}$ we get infinite number of terms of different order. 
\begin{eqnarray}
&&\frac{\partial f_{2}}{\partial t}+\frac{\bf{\xi}}{2m}\cdot
\frac{\partial f_{2}}{\partial {\bf{R}}}+\frac{2\bf{p}}{m}\cdot
\frac{\partial f_{2}}{\partial {\bf{r}}}-\frac{\partial
\phi}{\partial {\textbf{r}}}\cdot\frac{\partial f_{2}}{\partial
{\textbf{p}}}+ \sum_{j=1}^{\infty}(-1)^{j+1} C_{j}\hbar^{2j}\phi\left(\frac{\overleftarrow{\partial}^{2j+1}
}{\partial \textbf{r}^{2j+1}}
\frac{\overrightarrow{\partial}^{2j+1}}{\partial \textbf{p}^{2j+1}}\right)f_{2}=0 \label{v1}\end{eqnarray}
%AAAAAA
%\begin{eqnarray}
%&&\frac{\partial f_{2}}{\partial t}+\frac{\bf{\xi}}{2m}\cdot
%\frac{\partial f_{2}}{\partial {\bf{R}}}+\frac{2\bf{p}}{m}\cdot
%\frac{\partial f_{2}}{\partial {\bf{r}}}-\frac{\partial
%\phi}{\partial {\textbf{r}}}\cdot\frac{\partial f_{2}}{\partial
%{\textbf{p}}}+ C_{1}\hbar^{2}\phi\left(\frac{\overleftarrow{\partial}^{3}
%}{\partial \textbf{r}^{3}}
%\frac{\overrightarrow{\partial}^{3}}{\partial \textbf{p}^{3}}\right)f_{2}=0 \label{v1}\end{eqnarray}
where $\bf{R}=\frac{1}{2}(\bf{r_{1}}+\bf{r_{2}})$, $\bf{r}=\bf{r_{2}}-\bf{r_{1}}$, $\xi=(\bf{p_{1}}+\bf{p_{2}})$, $\bf{p}=\frac{1}{2}(\bf{p_{2}}-\bf{p_{1}})$. 
$\bf{r_{1}}$ and $\bf{p_{1}}$ are the position and momentum of the first particle and $\bf{r_{2}}$ and $\bf{p_{2}}$ are the position and momentum of the second particle respectively. In the normalized form the above equation in the steady state limit is
\begin{eqnarray}
&&\frac{\bf{\xi}}{2}\cdot
\frac{\partial f_{2}}{\partial {\bf{R}}}+2\textbf{p}\cdot
\frac{\partial f_{2}}{\partial {\bf{r}}}-\frac{\partial
\phi}{\partial {\textbf{r}}}\cdot\frac{\partial f_{2}}{\partial
{\textbf{p}}}+ \sum_{j=1}^{\infty}(-1)^{j+1} C_{j}\Lambda^{2j}\phi\left(\frac{\overleftarrow{\partial}^{2j+1}
}{\partial \textbf{r}^{2j+1}}
\frac{\overrightarrow{\partial}^{2j+1}}{\partial \textbf{p}^{2j+1}}\right)f_{2}=0\end{eqnarray}
$$C_{j}=1/{(2)}^{2j}(2j+1)!$$
with the following normalized variables.
$$t\sim\frac{t}{l\sqrt{m\beta}}$$
$${\bf{R}}\sim\frac{{\bf{R}}}{l},{\bf{r}}\sim\frac{{\bf{r}}}{l}$$
$${\bf{\xi}}\sim\frac{{\bf{\xi}}\sqrt{\beta}}{\sqrt{m}}, {\bf{p}}\sim\frac{{\bf{p}}\sqrt{\beta}}{\sqrt{m}}$$
 $$\Lambda =\sqrt{\frac{\hbar^{2}\beta}{ml^{2}}}$$
 $l$ is the length scale of the system and $\beta=1/k_{B}T$. $k_{B}$ is the Boltzmann constant and $T$ is the temperature of the system.
In this article we have focused on the first quantum correction of the  above equation. 
\begin{eqnarray}
&&\frac{\bf{\xi}}{2}\cdot
\frac{\partial f_{2}}{\partial {\bf{R}}}+2\textbf{p}\cdot
\frac{\partial f_{2}}{\partial {\bf{r}}}-\frac{\partial
\phi}{\partial {\textbf{r}}}\cdot\frac{\partial f_{2}}{\partial
{\textbf{p}}}+ C_{1}\Lambda^{2}\phi\left(\frac{\overleftarrow{\partial}^{3}
}{\partial \textbf{r}^{3}}
\frac{\overrightarrow{\partial}^{3}}{\partial \textbf{p}^{3}}\right)f_{2}=0 \label{q9}\end{eqnarray}

The subsequent terms are of the higher order of $\Lambda$ than the first two terms.
Therefore, we have omitted the next higher order terms of the series which is justified as long as the the value of $\Lambda$ is sufficiently small. Hence, this formalism can not explore system when the temperature is very low.

The
corresponding phase space distribution function obtained as the solution of eq.(\ref{q9}) is denoted by $f_2$
\begin{equation}f_{2}=e^{(-\frac{\xi^{2}}{4}+\mu_{1})}/[\exp(-a_{01}+a_{11}p^{2})-1]\label{v3}\end{equation}
where $a_{ij}$ are the functions of $r$ but not $p$.  $\mu_{1}$ is the chemical potential and normalized as $\beta\mu_{1}$.

$f_{2}$ can be expressed as
\begin{equation}f_{2}=e^{(-\frac{\xi^{2}}{4}+\mu_{1})}\sum_{n=1}^{\infty}\exp[n(a_{01}-a_{11}p^{2})]\label{v33}\end{equation}
We shall insert the above expression in eq.(\ref{q9}) as the trial solution and equate the coefficient of different power of the momentum to obtain a set of equation and solve them to obtain the coefficients $a_{01}$ and $a_{11}$. The process has been illustrated in the Appendix of \cite{kn:anirban1}.
The final form of the solution is given by
\begin{equation}f_{2}=e^{(-\frac{\xi^{2}}{4}+\mu_{1})}\sum_{n=1}^{\infty}e^{-n(\phi-\mu_{2}+p^{2})}K\label{q1}\end{equation}
where
$$K=1+\Lambda^{2}\left( \frac{n^{3}}{12}\left( \frac{\partial\phi}{\partial \textbf{r}}\right)^{2}-\frac{n^{2}}{4}\frac{\partial^{2}\phi}{\partial\textbf{r}^{2}}+\frac{n^{3}}{6}\left(\textbf{p}\cdot\frac{\partial}{\partial \textbf{r}}\right)^{2}\phi \right) $$
The first two terms of the series (n=1, 2) in eq.(\ref{q1}) are retained so that the correction due to the quantum statistics is included (to the lowest order) through the second term (n=2) of eq.(\ref{s1}).
\begin{eqnarray}f_{2}=e^{(-\frac{\xi^{2}}{4}+\mu_{1})}[e^{-(\phi-\mu_{2}+p^{2})}\left(1+\Lambda^{2}\left( \frac{1}{12}\left( \frac{\partial\phi}{\partial \textbf{r}}\right)^{2}-\frac{1}{4}\frac{\partial^{2}\phi}{\partial\textbf{r}^{2}}+\frac{1}{6}\left(\textbf{p}\cdot\frac{\partial}{\partial \textbf{r}}\right)^{2}\phi \right)\right)\nonumber
\\+e^{-2(\phi-\mu_{2}+p^{2})}\left(1+\Lambda^{2}\left( \frac{8}{12}\left( \frac{\partial\phi}{\partial \textbf{r}}\right)^{2}-\frac{4}{4}\frac{\partial^{2}\phi}{\partial\textbf{r}^{2}}+\frac{8}{6}\left(\textbf{p}\cdot\frac{\partial}{\partial \textbf{r}}\right)^{2}\phi \right)\right)]\label{s1}\end{eqnarray}

In first approximation only the classical term is retained in the non interacting limit.
\begin{equation}f_{2}=e^{-(-\mu_{1}-\mu_{2}+p^{2}+\frac{\xi^{2}}{4})}\end{equation}

The above distribution function can be integrated  over the whole phase space 
\begin{equation}N^{2}=e^{\mu_{1}+\mu_{2}}{\left(\frac{{V}}{\lambda^{3}}\right)}^2\end{equation} 
$$\lambda=\frac{h}{\sqrt{2\pi mkT}}, \rho =\frac{N}{V}$$

As in the relative coordinate system the effective mass of the particles are $m/2$, we have 
$e^{\mu_{2}}=2^{3/2}\rho \lambda^3$. Hence, $e^{\mu_{1}}=\rho \lambda^3/2^{3/2}$.
In the next approximation, we use the above value of  $e^{\mu_{1}}$ and find that $e^{\mu_{2}}=2^{3/2}\rho \lambda^3/(1+\rho \lambda^3)$. These values are inserted in the eq.(\ref{s1}) and keeping upto the first order correction terms of both kind and omitting the mixed term 
\begin{eqnarray}f_{2}=\frac{(\rho\lambda^{3})^2}{1+\rho\lambda^{3}}e^{-(\phi+p^{2}+\frac{\xi^{2}}{4})}\left(1+\Lambda^{2}\left( \frac{1}{12}\left( \frac{\partial\phi}{\partial \textbf{r}}\right)^{2}-\frac{1}{4}\frac{\partial^{2}\phi}{\partial\textbf{r}^{2}}+\frac{1}{6}\left(\textbf{p}\cdot\frac{\partial}{\partial \textbf{r}}\right)^{2}\phi
\right)\right)\nonumber
\\+\frac{2^{3/2}(\rho\lambda^{3})^3}{(1+\rho\lambda^{3})^2}e^{-2(\phi+p^{2})-\frac{\xi^{2}}{4}}\end{eqnarray}

It can be seen that if we include  more terms in addition to the first two terms of the series of eq.(\ref{q1}) then the new terms added to the above expression are of the higher order of $\rho\lambda^{3}$ than the first two terms. Hence, for small values of $\rho\lambda^{3}$ we may neglect those terms and we have restricted to the first two terms of the series.

Integrating over the whole phase space with the following normalization condition
\begin{equation}1=\frac{1}{V}\int SdV
\end{equation}
$S$ is identified as
\begin{equation}S=\frac{e^{-\phi}}{1+\rho\lambda^{3}}\left(1+\Lambda^{2}\left( \frac{{1}}{12}\left( \frac{\partial\phi}{\partial\textbf{r}}\right)^{2}-\frac{1}{6}\frac{\partial^{2}\phi}{\partial \textbf{r}^{2}}\right)\right)+\frac{\rho\lambda^{3}}{1+\rho\lambda^{3}}e^{-2\phi}\label{c5}\end{equation}
 If the quantum terms are omitted the classical value is
\begin{equation}S=e^{-\phi}\end{equation}

$S$ as given by eq.(\ref{c5}),  may be inserted in eq.(\ref{c6}) to calculate the second virial coefficient and that can be expressed in the following form 
\begin{equation}B=B_{cl}+B_{1}+B_{ex}
\end{equation}
$B_{cl}$ is the classical part and $B_{1}$ is the contribution of the differential quotients of the potential and proportional to $\hbar^{2}$. $B_{ex}$ is the contribution of the quantum statistics and proportional to $\hbar^{3}$. Terms proportional to $\hbar^{4}$ and higher orders are omitted.

 \begin{equation}B_{cl}=-2\pi N\int_{0}^{\infty}(e^{-\phi}-1)r^{2}dr\label{c1}\end{equation}
\begin{equation}B_{1}=2\pi N\frac{\hbar^{2}}{12mk^{3}T^{3}}\int_{0}^{\infty} e^{-\phi}\left(\frac{d\phi}{dr}\right)^{2}r^{2}dr\label{c2}\end{equation}
\begin{equation}B_{ex}=-2\pi N \rho\lambda^{3}\int_{0}^{\infty} (e^{-\phi}+e^{-2\phi}-2)r^{2}dr\label{c3}\end{equation}

\section{Discussion}
First of all, this result is not in complete  agreement with that obtained by J. de Boer in 1949 \cite{kn:boer}. The expressions of the first two parts ($B_{cl}, B_{1}$) of the second virial coefficient are identical with that of the work done by J. de Boer. But the last part ($B_{ex}$) arising out of the inclusion of the indistinguishibility differs from each other. In our case $B_{ex}$ depends on the radius of the gas molecules in the expression. On the contrary, the expression of J. de Boer does not have radius of the molecules. In other way we may find that there is an additional fractional volume factor (volume of the all gas molecules/ volume of the container)  which is missing in the de Boer result. The reason behind the difference lies in the distribution functions employed to the respective cases. The distribution functions in these two cases are identical in the first part upto the $\hbar^2$ order term where the first order quantum correction to the classical results have been incorporated. But they differ when the first order correction due to the indistinguishibility factor comes into play. In this article we have been able to incorporate both the quantum effects in a single distribution function by using the formalism to determine the solution of the Wigner equation. In De Boer case there is no single distribution function which incorporates both the corrections. He derived the two different correction terms separately and superposed them two obtain the final virial coefficient.  He applied  the distribution function which was originally proposed by Kirkwood \cite{kn:kirk} in 1931 and was applied to this case by Uhlenbeck and Beth \cite{kn:uhl} in 1936. By starting from an equation obtained by Bloch \cite{kn:blo}, Kirkwood established a recursion relation to calculate the distribution function which makes his technique simpler, especially for higher order approximations, than the method employed by Wigner \cite{kn:wigner}. In addition the correction due to the symmetry restrictions to be placed on the wave functions according to the Fermi-Dirac or Bose-Einstein was obtained by Kirkwood. This symmetry restriction was originally neglected by Wigner. Consequently, the distribution Wigner obtained in the semi classical limit does not take care of the symmetry effect and it can only be expanded in even power of $h$ which is in contrast to the Kirkwood's result which contains terms of both even and odd power of $h$. In \cite{kn:anirban1} we have overcome this limitation of Wigner's approach by picking the Bose-Einstein distribution as the unperturbed solution of the Wigner equation instead of the Maxwell-Boltzmann distribution. This choice enables to incorporate the flavor of the quantum statistics in the calculation. In addition, we have been able to obtain recursion relation of higher order coefficients even in the Wigner's original formalism.

The above integrals (eq.(\ref{c1})-eq.(\ref{c3})) is evaluated with the Lenard-Jones potential as $\phi$. The $\phi$ is given by\cite{kn:boer1}
 \begin{equation}\phi=4\epsilon\left(\frac{1}{R^{12}}-\frac{1}{R^{6}}\right)\end{equation}
where $R=r/\sigma$.
We have chosen the values of N (6.023$\times 10^{23}$), $\sigma$ ($2.597\times 10^{-8}$cm), $\theta$ (2.67) and $\epsilon$ ($9.49\times 10^{-18}$ergs) from the data provided by J. de Boer in \cite{kn:boer1}. Finally, we obtain
\begin{equation}B_{cl}=\frac{2\pi N\sigma^3}{3}\left(\frac{4\epsilon}{kT}\right)^{1/4}\sum_{0}^{\infty}c_{i} \left(\frac{4\epsilon}{kT}\right)^{i/2}\end{equation}

where $c_{i}=-\frac{1}{4i!}\Gamma\left(-\frac{1}{4}+\frac{i}{2}\right)$
\begin{equation}B_{1}=\frac{2\pi N\sigma h^2}{3m\epsilon}\left(\frac{4\epsilon}{kT}\right)^{13/12}\sum_{0}^{\infty}d_{i} \left(\frac{4\epsilon}{kT}\right)^{i/2}\end{equation}

where $d_{i}=\frac{36i-11}{768\pi^{2}i!}\Gamma\left(-\frac{1}{12}+\frac{i}{2}\right)$
\begin{equation}B_{ex}=\frac{2\pi N\sigma^3}{3}\rho \lambda^3 \left[\left(\frac{4\epsilon}{kT}\right)^{1/4}\sum_{0}^{\infty}c_{i} \left(\frac{4\epsilon}{kT}\right)^{i/2}+\left(\frac{8\epsilon}{kT}\right)^{1/4}\sum_{0}^{\infty}c_{i} \left(\frac{8\epsilon}{kT}\right)^{i/2}\right]\end{equation}

The expression of $B_{ex}$ obtained by J. de Boer is given by
\begin{equation}B_{ex}=N\frac{\lambda^3}{2^{5/2}}\end{equation}

\section{Conclusion}
At the end, it is concluded that the formalism developed in \cite{kn:anirban1} to determine the distribution function of the non interacting Bosons in the presence of external potential has been successfully applied in the case of interacting Bosons to determine the quantum correction to the second virial coeeficient of the Bose gas interacting pairwise via Lenard-Jones potential. The single  distribution function containing both the quantum effects is sufficient to produce the complete result. On the contrary, in the previous attempt by J. de Boer the two quantum corrections were calculated separately and finally added them to get the complete result. Consequently,  the part of the result originated  from the indistinguishibility factor differs from the previous result obtained by J. de Boer.

 It is observed that in addition to the density and temperature of the system, the differential quotients of the potential play \cite{kn:wigner,kn:uhl} a vital part in determining the strength of quantum corrections. The process of including the higher order corrections is straight forward. For example, we could include the $\Lambda^{4}$ order correction by including the $\Lambda^{4}$ term \cite{kn:imam}  in eq.(\ref{v1}). We are interested only in the lowest order corrections of two distinct effects - the first one is related to the inclusion of the terms containing the higher derivatives of the potential in  addition to the usual classical term of the Boltzmann equation and the second one is the exchange effect. We have omitted the subsequent higher order corrections. We have also omitted the mixed terms containing both the effects. 

The mathematical procedure begins from the equation of the two particle distribution function of a system of molecules available  in the quantum mechanical generalization of the Boltzmann equation. The approximation is the binary encounter approximation which needs the density of the system to be sufficiently small so that the probability of encounters involving more than two particles are 
very low. We have identified the ratio of the thermal De Broglie length of a molecule and the length scale of the system as the expansion parameter ($\Lambda$). The temperature should be sufficiently large to make the expansion parameter sufficiently small to achieve the convergence of the problem. Hence, the formalism is limited to the small values of $\Lambda$ and sufficiently low density systems.

The data that support the findings of this study are available from the corresponding author upon reasonable request.

\end{document}